\documentclass[preprintnumbers,amsmath,amssymb,prd,12pt,floatfix,superscriptaddress,nofootinbib]{revtex4}
\usepackage{graphicx}
\usepackage{epsfig}
\usepackage{bm}
\usepackage{amsfonts}

\newcommand{\la}{\lambda}

\def\be {\begin{equation}}
\def\ee {\end{equation}}
\def\ba {\begin{eqnarray}}
\def\ea {\end{eqnarray}}
\newcommand{\bear}{\begin{eqnarray}}

\newcommand{\eear}{\end{eqnarray}}

\newbox\pippobox

\def\rl{\rho_\Lambda}

\def\6{\partial}

\def\la{\Lambda}
\def\a{\alpha}

\def\ol{\Omega_\Lambda}

\def\sq
\def\a{\alpha}

\def\bx{{\bf x}}

\begin{document}

\title{\bf Holographic dark energy with varying gravitational constant in Ho\v{r}ava-Lifshitz
 cosmology\footnote{\it M. R. Setare dedicated this paper to the 70 year Jubilee of
Professor Farhad Ardalan }}

\author{M. R. Setare}
\email{rezakord@ipm.ir} \affiliation{Department of Physics, University of
Kurdistan, Pasdaran Ave., Sanandaj, Iran
}

\author{Mubasher Jamil}
\email{mjamil@camp.nust.edu.pk} \affiliation{Center for Advanced
Mathematics and Physics, National University of Sciences and
Technology, Rawalpindi, 46000, Pakistan}

\begin{abstract}
\textbf{Abstract:} We investigate the holographic dark energy
scenario with a varying gravitational constant in a flat
background in the context of Ho\v{r}ava-Lifshitz gravity. We
extract the exact differential equation determining the evolution
of the dark energy density parameter,
 which includes $G$ variation term. Also we discuss  a cosmological
  implication of our work by
  evaluating the dark energy equation of state for low redshifts
  containing varying $G$ corrections.
\end{abstract}

 \maketitle

\section{Introduction}
Observational data indicates that our universe is currently under
accelerating expansion \cite{{1},{111},2}. This acceleration implies
that if Einstein's theory of gravity is reliable on cosmological
scales, then our universe is dominated by a mysterious form of
energy. This unknown energy component possesses some strange
features, for example it is not clustered on large length scales and
its pressure must be negative so that can drive the current
acceleration of the universe. Since the fundamental theory of nature
that could explain the microscopic physics of DE is unknown at
present, phenomenologists take delight in constructing
various models based on its macroscopic behavior. \\
Despite the lack of a quantum theory of gravity, we can still make
some attempts to probe the nature of dark energy according to some
principles of quantum gravity. An interesting attempt in this
direction is the so-called ``holographic dark energy'' (HDE)
proposal \cite{Cohen:1998zx,Hsu:2004ri,Li:2004rb,holoext}. The HDE
is defined by \be \rho_\Lambda=\frac{3c^2}{8\pi GL^2}.\ee Here $c^2$
is a holographic parameter of order unity and $G$ is a gravitational
`constant'. In this chapter, we shall treat $G$ as a variable. The
HDE paradigm has been constructed in the light of holographic
principle of quantum gravity \cite{holoprin}, and thus it presents
some interesting features of an underlying theory of dark energy.
Furthermore, it may simultaneously provide a solution to the
coincidence problem, i.e why matter and dark energy densities are
comparable today although they obey completely different equations
of motion \cite{Li:2004rb}. The holographic dark energy model has
been extended to include the spatial curvature contribution
\cite{nonflat}. Lastly, it has been tested and constrained by
various astronomical observations \cite{obs3}.\\
 Recently, a
power-counting renormalizable, ultra-violet (UV) complete theory of
gravity was proposed by Ho\v{r}ava in \cite{hor2,hor1,hor3,hor4}.
Although presenting an infrared (IR) fixed point, namely General
Relativity, in the  UV the theory possesses a fixed point with an
anisotropic, Lifshitz scaling between time and space of the form
$\bx\to\ell~\bx$, $t\to\ell^z~t$, where $\ell$, $z$, $\bx$ and $t$
are the scaling factor, dynamical critical exponent, spatial
coordinates and temporal coordinate, respectively.\\
 Due to these
novel features, there has been a large amount of effort in examining
and extending the properties of the theory itself
\cite{Volovik:2009av,Orlando:2009en,Nishioka:2009iq,Konoplya:2009ig,Charmousis:2009tc,Li:2009bg,Visser:2009fg,Sotiriou:2009bx,Chen:2009bu,Chen:2009ka,Shu:2009gc,Bogdanos:2009uj,Kluson:2009rk}.
Additionally, application of Ho\v{r}ava-Lifshitz gravity as a
cosmological framework gives rise to Ho\v{r}ava-Lifshitz cosmology,
which proves to lead to interesting behavior
\cite{Calcagni:2009ar,Kiritsis:2009sh}. In particular, one can
examine specific solution subclasses
\cite{Lu:2009em,Nastase:2009nk,Minamitsuji:2009ii}, the perturbation
spectrum
\cite{Gao:2009bx,Chen:2009jr,Gao:2009ht,Wang:2009yz,Kobayashi:2009hh},
the gravitational wave production
\cite{Takahashi:2009wc,Koh:2009cy}, the matter bounce
\cite{Brandenberger:2009yt,Brandenberger:2009ic,Cai:2009in}, the
black hole properties \cite{Danielsson:2009gi,Kehagias:2009is,
Mann:2009yx,Bertoldi:2009vn,Castillo:2009ci,BottaCantcheff:2009mp},
the dark energy phenomenology
\cite{Saridakis:2009bv,Wang:2009rw,Appignani:2009dy,Setare:2009vm},
the astrophysical phenomenology \cite{Kim:2009dq,Harko:2009qr} etc.
However, despite this extended research, there are still many
ambiguities if Ho\v{r}ava-Lifshitz gravity is reliable and capable
of a successful description of the gravitational background of our
world, as well as of the cosmological behavior of the universe
\cite{Charmousis:2009tc,Sotiriou:2009bx,Bogdanos:2009uj}.\\
In the present work we are interested to study the
 Holographic dark energy in framework of Ho\v{r}ava-Lifshitz
 gravity. We extend our analysis with considering the time
variable Newton's constant $G$.  Until now, in most the investigated
dark energy models a constant Newton's ``constant'' $G$ has been
considered. However, there are significant indications that $G$ can
by varying, being a function of time or equivalently of the scale
factor \cite{G4com}. In particular, observations of Hulse-Taylor
binary pulsar \cite{Damour,kogan}, helio-seismological data
\cite{guenther} and astro-seismological data from the pulsating
white dwarf star G117-B15A \cite{Biesiada} lead to
$\left|\dot{G}/G\right| \lessapprox 4.10 \times 10^{-11} yr^{-1}$,
for $z\lesssim3.5$ \cite{ray1}. Additionally, a varying $G$ has some
theoretical advantages too, alleviating the dark matter problem
\cite{gol}, the cosmic coincidence problem \cite{jamil} and the
discrepancies in Hubble parameter value \cite{ber}.\\
The plan of the paper is as follows: In the second section we shall present a brief overview of Ho\v{r}ava-Lifshitz cosmology. In third section we construct the holographic dark energy with varying gravitational constant and extract the differential equation that determines the evolution of dark energy parameter. In section four, we use these expressions in order to calculate the corrections to the dark energy equation of state for low redshift. Finally in section five, we briefly discuss our results.

\section{Ho\v{r}ava-Lifshitz cosmology}
\label{model}

\subsection{Dark-matter field formulation}

 We begin with a brief review of Ho\v{r}ava-Lifshitz
gravity. The dynamical variables are the lapse and shift
functions, $N$ and $N_i$ respectively, and the spatial metric
$g_{ij}$ (roman letters indicate spatial indices). In terms of
these fields the full metric is
\begin{eqnarray}
ds^2 = - N^2 dt^2 + g_{ij} (dx^i + N^i dt ) ( dx^j + N^j dt ) ,
\end{eqnarray} %%
where indices are raised and lowered using $g_{ij}$. The scaling
transformation of the coordinates reads ($z=3$):
\begin{eqnarray}
 t \rightarrow l^3 t~~~{\rm and}\ \ x^i \rightarrow l x^i~.
\end{eqnarray}

\subsubsection{Detailed Balance}

Decomposing the gravitational action into a kinetic and a
potential part as $S_g = \int dt d^3x \sqrt{g} N ({\cal L}_K+{\cal
L}_V)$, and under the assumption of detailed balance \cite{hor3}
(the extension beyond detail balance will be performed later on),
which apart form reducing the possible terms in the Lagrangian it
allows for a quantum inheritance principle \cite{hor2} (the $D+1$
dimensional theory acquires the renormalization properties of the
D-dimensional one),
 the full action of Ho\v{r}ava-Lifshitz gravity is given by
\begin{eqnarray}
S_g =  \int dt d^3x \sqrt{g} N \left\{ \frac{2}{\kappa^2}
(K_{ij}K^{ij} - \lambda K^2)- \ \ \ \ \ \ \ \ \ \ \ \ \ \ \ \ \  \right. \nonumber \\
\left.
 - \frac{\kappa^2}{2 w^4} C_{ij}C^{ij}
 + \frac{\kappa^2 \mu}{2 w^2}
\frac{\epsilon^{ijk}}{\sqrt{g}} R_{il} \nabla_j R^l_k -
\frac{\kappa^2 \mu^2}{8} R_{ij} R^{ij}+
     \right. \nonumber \\
\left.    + \frac{\kappa^2 \mu^2}{8(1 - 3 \lambda)} \left[ \frac{1
- 4 \lambda}{4} R^2 + \Lambda  R - 3 \Lambda ^2 \right] \right\},
\end{eqnarray}
where
\begin{eqnarray}
K_{ij} = \frac{1}{2N} \left( {\dot{g_{ij}}} - \nabla_i N_j -
\nabla_j N_i \right) \, ,
\end{eqnarray}
is the extrinsic curvature and
\begin{eqnarray} C^{ij} \, = \, \frac{\epsilon^{ijk}}{\sqrt{g}} \nabla_k
\bigl( R^j_i - \frac{1}{4} R \delta^j_i \bigr),
\end{eqnarray}
is the Cotton tensor, and the covariant derivatives are defined with
respect to the spatial metric $g_{ij}$. $\epsilon^{ijk}$ is the
totally antisymmetric unit tensor, $\lambda$ is a dimensionless
constant and $\Lambda $ is a negative constant which is related to
the cosmological constant in the IR limit. Finally, the variables
$\kappa$, $w$ and $\mu$ are constants with mass dimensions $-1$,
$0$ and $1$, respectively.

In order to add the dark-matter content in a universe governed by
Ho\v{r}ava gravity, a scalar field is introduced
\cite{Calcagni:2009ar,Kiritsis:2009sh}, with action:
\begin{eqnarray}
S_m\equiv S_\phi = \int dtd^3x \sqrt{g} N \left[
\frac{3\lambda-1}{4}\frac{\dot\phi^2}{N^2}
+m_1m_2\phi\nabla^2\phi-\right.\nonumber\\
\left.-\frac{1}{2}m_2^2\phi\nabla^4\phi +
\frac{1}{2}m_3^2\phi\nabla^6\phi -V(\phi)\right],\ \ \ \ \
\end{eqnarray}
where $V(\phi)$ acts as a potential term and $m_i$ are constants.
Although one could just follow a hydrodynamical approximation and
introduce straightaway the density and pressure of a matter fluid
\cite{Sotiriou:2009bx}, the field approach is more robust,
especially if one desires to perform a phase-space analysis.

Now, in order to focus on cosmological frameworks, we impose the
so called projectability condition \cite{Charmousis:2009tc} and
use an FRW metric,
\begin{eqnarray}
N=1~,~~g_{ij}=a^2(t)\gamma_{ij}~,~~N^i=0~,
\end{eqnarray}
with
\begin{eqnarray}
\gamma_{ij}dx^idx^j=\frac{dr^2}{1-kr^2}+r^2d\Omega_2^2~,
\end{eqnarray}
where $k=-1,0,1$ correspond to open, flat, and closed universe
respectively. In addition, we assume that the scalar field is
homogenous, i.e $\phi\equiv\phi(t)$. By varying $N$ and $g_{ij}$,
we obtain the equations of motion:
\begin{eqnarray}\label{Fr1}
H^2 &=&
\frac{\kappa^2}{6(3\la-1)}\left[\frac{3\la-1}{4}\,\dot\phi^2
+V(\phi)\right]+\nonumber\\
&+&\frac{\kappa^2}{6(3\la-1)}\left[
-\frac{3\kappa^2\mu^2k^2}{8(3\lambda-1)a^4}
-\frac{3\kappa^2\mu^2\Lambda ^2}{8(3\lambda-1)}
 \right]+\nonumber\\
 &+&\frac{\kappa^4\mu^2\Lambda k}{8(3\lambda-1)^2a^2} \ ,
\end{eqnarray}
\begin{eqnarray}\label{Fr2}
\dot{H}+\frac{3}{2}H^2 &=&
-\frac{\kappa^2}{4(3\la-1)}\left[\frac{3\la-1}{4}\,\dot\phi^2
-V(\phi)\right]-\nonumber\\
&-&\frac{\kappa^2}{4(3\la-1)}\left[-\frac{\kappa^2\mu^2k^2}{8(3\lambda-1)a^4}
+\frac{3\kappa^2\mu^2\Lambda ^2}{8(3\lambda-1)}
 \right]+\nonumber\\
 &+&\frac{\kappa^4\mu^2\Lambda k}{16(3\lambda-1)^2a^2}\ ,
\end{eqnarray}
where we have defined the Hubble parameter as $H\equiv\frac{\dot
a}{a}$. Finally, the equation of motion for the scalar field
reads:
\begin{eqnarray}\label{phidott}
&&\ddot\phi+3H\dot\phi+\frac{2}{3\lambda-1}\frac{dV(\phi)}{d\phi}=0.
\end{eqnarray}

At this stage we can define the energy density and pressure for
the scalar field responsible for the matter content of the
Ho\v{r}ava-Lifshitz universe:
\begin{eqnarray}
&&\rho_m\equiv \rho_\phi=\frac{3\la-1}{4}\,\dot\phi^2
+V(\phi)\label{rhom}\\
&&p_m\equiv p_\phi=\frac{3\la-1}{4}\,\dot\phi^2
-V(\phi).\label{pressurem}
\end{eqnarray}
Concerning the dark-energy sector we can define
\begin{equation}\label{rhoDE}
\rho_\Lambda\equiv -\frac{3\kappa^2\mu^2k^2}{8(3\lambda-1)a^4}
-\frac{3\kappa^2\mu^2\Lambda ^2}{8(3\lambda-1)}
\end{equation}
\begin{equation}
\label{pDE} p_\Lambda\equiv
-\frac{\kappa^2\mu^2k^2}{8(3\lambda-1)a^4}
+\frac{3\kappa^2\mu^2\Lambda ^2}{8(3\lambda-1)}.
\end{equation}
The term proportional to $a^{-4}$ is the usual ``dark radiation
term'', present in Ho\v{r}ava-Lifshitz cosmology
\cite{Calcagni:2009ar,Kiritsis:2009sh}. Finally, the constant term
is just the explicit (negative) cosmological constant. Therefore,
in expressions (\ref{rhoDE}),(\ref{pDE}) we have defined the
energy density and pressure for the effective dark energy, which
incorporates the aforementioned contributions.

Using the above definitions, we can re-write the Friedmann
equations (\ref{Fr1}),(\ref{Fr2}) in the standard form:
\begin{eqnarray}
\label{Fr1b} H^2 &=&
\frac{\kappa^2}{6(3\lambda-1)}\Big[\rho_m+\rho_\Lambda\Big]+ \frac{\beta k}{a^2},\\
\label{Fr2b} \dot{H}+\frac{3}{2}H^2 &=&
-\frac{\kappa^2}{4(3\lambda-1)}\Big[p_m+p_\Lambda
 \Big]+ \frac{\beta k}{2a^2}.
\end{eqnarray}
In these relations we have defined $\kappa^2=8\pi G$ and
$\beta\equiv\frac{\kappa^4\mu^2\Lambda }{8(3\lambda-1)^2}$, which
is the coefficient of the curvature term. Additionally, we could
also define an effective Newton's constant and an effective light
speed \cite{Calcagni:2009ar,Kiritsis:2009sh}, but we prefer to
keep $\frac{\kappa^2}{6(3\la-1)}$ in the expressions, just to make
clear the origin of these terms in Ho\v{r}ava-Lifshitz cosmology.
Finally, note that using (\ref{phidott}) it is straightforward to
see that the aforementioned dark matter and dark energy quantities
verify the standard evolution equations:
\begin{eqnarray}\label{phidot2}
&&\dot{\rho}_m+3H(\rho_m+p_m)=0,\\
\label{sdot2} &&\dot{\rho}_\Lambda+3H(\rho_\Lambda+p_\Lambda)=0.
\end{eqnarray}

\section{Holographic Dark Energy with varying gravitational constant in a flat background}
\label{model}
Let us construct holographic dark energy scenario allowing for a
varying Newton's constant $G$. The space-time geometry will be a
flat Robertson-Walker:
\begin{equation}\label{met}
ds^{2}=-dt^{2}+a(t)^{2}(dr^{2}+r^{2}d\Omega_2^{2}),
\end{equation}
with $a(t)$ the scale factor and $t$ the comoving time. As usual,
the first Friedmann equation reads:
\begin{eqnarray}
\label{Fr1b} H^2 &=& \frac{8\pi G}{6(3\lambda-1)}\rho \label{Fr2b}
\end{eqnarray}
with $H$ the Hubble parameter, $\rho=\rho_m+\rho_\Lambda $,
$\rho_m=\rho_{m0}a^{-3}$, where $\rho_m$ and $\rl$ stand
respectively for  matter and dark energy densities and the index $0$
marks the present value of a quantity. Furthermore, we will use the
density parameter $ \Omega_\Lambda\equiv\frac{8\pi G}{3H^2}\rl$, which,
imposing explicitly the holographic nature of dark energy according
to relation , becomes
\begin{eqnarray}
 \label{OmegaL2}
\Omega_{\Lambda}=\frac{c^2}{H^2L^2}.
\end{eqnarray}
 Finally,
in the case of a flat universe, the best choice for the definition
of $L$ is to identify it with the future event horizon
\cite{Li:2004rb}, that is $L\equiv R_ h(a)$ with
\begin{equation}
 R_ h(a)=a\int_t^\infty{dt'\over
a(t')}=a\int_a^\infty{da'\over Ha'^2}~.\label{eh}
\end{equation}

In the following we will use $\ln a$ as an independent variable.
Thus, denoting by dot the time-derivative and by prime the
derivative with respect to $\ln a$, for every quantity $F$ we
acquire $\dot{F}=F'H$. Differentiating (\ref{OmegaL2}) using
(\ref{eh}), and observing that $\dot{R}_h=HR_h-1$, we obtain:
\begin{equation}\label{OmegaLdif}
\frac{\ol'}{\ol^2}=\frac{2}{\ol}\Big[-1-\frac{\dot{H}}{H^2}+\frac{\sqrt{\ol}}{c}\Big].
\end{equation}
Until now, the varying behavior of $G$ has not become manifested.
However, the next step is to eliminate $\dot{H}$. This can be
obtained by differentiating Friedman equation, leading to
\begin{equation}
\dot{H}=\frac{2\pi}{3(3\lambda-1)H}(\dot G\rho+G\dot\rho),
\end{equation}
where $G$ is considered to be a function of $t$. Using \be \dot
G=G^\prime H,\ee and the energy conservation equation  \be
\dot\rho=-3H(1+\omega)\rho,\ee where $\omega$ is \be
 \omega=\frac{\omega_\Lambda\rho_\Lambda}{\rho}=\frac{\omega_\Lambda\Omega_\Lambda}{2(3\lambda-1)}.
\ee The EoS parameter for HDE is given by \be
\omega_\Lambda=-\Big(\frac{1}{3}+\frac{2\sqrt{\Omega_\Lambda}}{3c}\Big)
\ee Eq. (26) becomes \be \frac{2\dot
H}{H^2}=\Delta_G-3(1+\omega),\ee where $\Delta_G\equiv G^\prime/G $
is a dimensionless number.   \be \frac{2\dot H}{H^2}=\Delta_G-3\Big[
1-\frac{\Omega_\Lambda}{2(3\lambda-1)}
\Big(\frac{1}{3}+\frac{2\sqrt{\Omega_\Lambda}}{3c}\Big) \Big], \ee
Using (32) in (25) we get finally \be
\frac{\Omega_\Lambda^\prime}{\Omega_\Lambda^2}=\frac{2}{\Omega_\Lambda}\Big[2-\frac{\Delta_G}{2}+\frac{\sqrt{\Omega_\Lambda}}{2c}
+\frac{\Omega_\Lambda}{4(3\lambda-1)}+\frac{\Omega_\Lambda^{3/2}}{2c(3\lambda-1)}\Big]
\ee

\section{Cosmological implications}

What we are interested in most is the prediction about the equation
of state at the present time. Since we have extracted the
expressions for $\Omega_\Lambda'$, we can calculate $w(z^\prime)$ for small
redshifts $z^\prime$, performing the standard expansions of the
literature. In particular, since $\rho_\la\sim a^{-3(1+w)}$ we
acquire after expanding $\rho_\la$ as
\begin{equation}
\ln\rho_\la =\ln \rho_\la^0+{d\ln\rho_\la \over d\ln a} \ln a
+\frac{1}{2} {d^2\ln\rho_\la \over d(\ln a)^2}(\ln a)^2+\dots,
\end{equation}
 where the
derivatives are taken at the present time $a_0=1$ (and thus at
$\ol=\Omega_{\la}^0$). Then, $w(\ln a)$ is given as
\begin{equation}
w(\ln a)=-1-{1\over 3}\left[{d\ln\rho_\la \over d\ln a}
+\frac{1}{2} {d^2\ln\rho_\la \over d(\ln a)^2}\ln a\right],
\end{equation}
 up to second order.
 In addition, we can straightforwardly calculate
$w(z^\prime)$, replacing $\ln a=-\ln(1+z^\prime)\simeq -z^\prime$,
which is valid for small redshifts, defining
\begin{equation}
w(z)=-1-{1\over 3}\left({d\ln\rho_\la \over d\ln a}\right)+
\frac{1}{6} \left[{d^2\ln\rho_\la \over d(\ln
a)^2}\right]\,z^\prime\equiv w_0+w_1z^\prime.
\end{equation}
 Since $\rho_\Lambda =3H^2 \Omega_\Lambda/(8\pi G)= \Omega_\Lambda
\rho_m/\Omega_m=\rho_{m0}\Omega_\Lambda/(1 -\Omega_\Lambda) a^{-3}$, the
derivatives are easily computed using  the obtained expressions for
$\ol'$. Hence we get \ba
\omega_0&=&-\frac{2}{3}-\frac{\Omega_\Lambda^\prime}{\Omega_\Lambda(1-\Omega_\Lambda)},\\
\omega_1&=&\frac{1}{6}\Big[
\frac{\Omega_\Lambda^{\prime\prime}}{\Omega_\Lambda(1-\Omega_\Lambda)}-\frac{\Omega_\Lambda^{\prime2}}{(1-\Omega_\Lambda)^2}(1+2\Omega_\Lambda+2\Omega_\Lambda^2)
\Big]. \ea Using Eq. (33) in (37) and (38), we get \ba
\omega_0&=&-\frac{2}{3}-\frac{1}{1-\Omega_\Lambda}\Big[
4-\Delta_G+\frac{\sqrt{\Omega_\Lambda}}{c}+\frac{\Omega_\Lambda}{2(3\lambda-1)}+\frac{\Omega_\Lambda^{3/2}}{c(3\lambda-1)}
\Big].\ea \ba
 \omega_1&=&\frac{1}{6}\Big[ \frac{1}{1-\Omega_\Lambda}\Big(
4-\Delta_G+\frac{\sqrt{\Omega_\Lambda}}{c}+\frac{\Omega_\Lambda}{2(3\lambda-1)}+\frac{\Omega_\Lambda^{3/2}}{c(3\lambda-1)}
\Big) \nonumber\\&\;&\times\Big(
4-\Delta_G+\frac{3\sqrt{\Omega_\Lambda}}{2c}+\frac{\Omega_\Lambda}{3\lambda-1}+\frac{5\Omega_\Lambda^{3/2}}{2c(3\lambda-1)}
\Big)\nonumber\\&\;&-\frac{\Omega_\Lambda^2}{(1-\Omega_\Lambda)^2}(1+2\Omega_\Lambda+2\Omega_\Lambda^2)\nonumber\\&\;&
\times\Big(4-\Delta_G+\frac{\sqrt{\Omega_\Lambda}}{c}+\frac{\Omega_\Lambda}{2(3\lambda-1)}+\frac{\Omega_\Lambda^{3/2}}{c(3\lambda-1)}
\Big)\Big]. \ea Eqs. (39) and (40) together determine the linear
equation of state of dark energy. In order to choose their numerical
values and fit them to the observational data, we require a suitable
value of $\lambda$. None of the numerical values i.e.
$\lambda=1,1/3$ or $\infty$ gives a good estimate. Hence we think
that the appropriate value of $\lambda$ should be deduced from the
observational data.

\section{Conclusions}
\label{Conclusions}
Astrophysical observations suggest that dark energy state parameter must be variable
and changing over cosmic time. In this connection, a proposed dark energy candidate
is the holographic dark energy. The HDE naturally represents a variable form of dark
 energy by involving a parameter $c$ in it. The choice $c>1$ leads to the dark energy
 behaving as phantom while for $c>1$, in a situation without any other component of
 energy, the spacetime is not de Sitter. For various reasons, people choose $c=1$. \\
In this paper, we have analyzed the holographic dark energy with varying gravitational
 constant in the framework of Ho\v{r}ava gravity. This work is motivated from various
  astrophysical observations that point out
  to the notion of variable constants.  Although the variations are negligibly small,
  nevertheless these could be significant in the overall cosmological evolution. It is
   shown here, the equation of state of dark energy, for lower redshift $z\sim1$
   approximation, is modified due to contribution of varying $G$. \\

\end{document}